\documentclass[journal=jacsat,manuscript=article]{achemso}

\usepackage{chemformula} 
\usepackage[T1]{fontenc} 

\mciteErrorOnUnknownfalse

\author{Vishnu Sudarsanan}
\affiliation[CUTN]
{Department of Physics, Central University of Tamil Nadu, Thiruvarur 610005, India}
\alsoaffiliation[SCANMAT]{Simulation Center for Atomic and Nanoscale MATerials, Central University of Tamil Nadu, Thiruvarur, Tamil Nadu, 610005, India.}
\author{Anu Maria Augustine}
\affiliation[CUTN]
{Department of Physics, Central University of Tamil Nadu, Thiruvarur 610005, India}
\alsoaffiliation[SCANMAT]{Simulation Center for Atomic and Nanoscale MATerials, Central University of Tamil Nadu, Thiruvarur, Tamil Nadu, 610005, India.}
\author{P Ravindran}
\email{raviphy@cutn.ac.in}
\affiliation[CUTN]
{Department of Physics, Central University of Tamil Nadu, Thiruvarur 610005, India}
\alsoaffiliation[SCANMAT]{Simulation Center for Atomic and Nanoscale MATerials, Central University of Tamil Nadu, Thiruvarur, Tamil Nadu, 610005, India.}
\alsoaffiliation[Norway]
{Center for Materials Science and Nanotechnology and Department of Chemistry, University of Oslo, Box 1033 Blindern, N-0315 Oslo, Norway.}

\title[An \textsf{achemso} demo]
  {Investigation of Electronic Structure and Electrochemical Properties of Na\textsubscript{2}MnSiO\textsubscript{4} as a Cathode Material for Na-ion Batteries}

\abbreviations{IR,NMR,UV}

\begin{document}

\begin{tocentry}
    \includegraphics[scale=0.68]{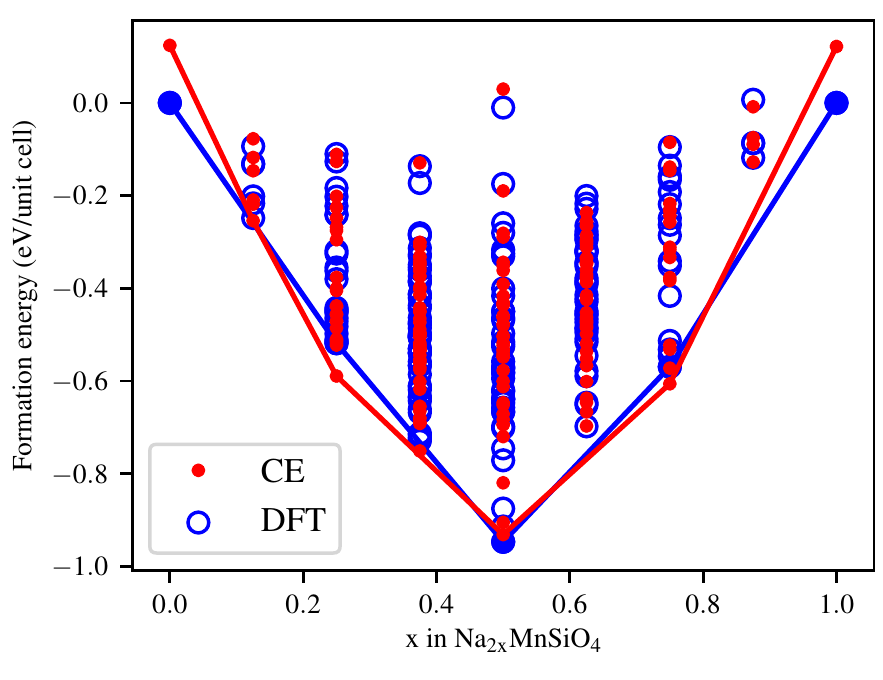}
    \label{graph_abstract}
\end{tocentry}

\begin{abstract}
  The polyanionic compound Na\textsubscript{2}MnSiO\textsubscript{4} is regarded as one of the promising cathode materials for Na-ion batteries due to  good specific capacity with its attractive prospect of utilization of two electrons in the redox processes. So, in this study, we have performed the thermodynamic and electronic structure analysis of  Na\textsubscript{2}MnSiO\textsubscript{4} using first principles density functional theory calculations. The intermediate ground state configurations for Na\textsubscript{2}MnSiO\textsubscript{4} of Na de-intercalation were found using the cluster expansion method and are used to obtain the $0~K$ voltage profile as a function of Na concentration. This material shows an average voltage of 4.2~V and the finite temperature analysis at $300~K$ using Monte Carlo simulations indicates that this material undergoes two phase mixing when desodiate beyond 1.5~Na/f.u.  The chemical bonding interactions between the constituents were analyzed using various bond analysis tools. The involvement of oxygen in the redox reaction apart from the transition metal is identified using the Bader charge analysis. Relevant Na diffusion pathways and their corresponding calculated energy barriers are compared with the partially Fe substituted Na\textsubscript{2}MnSiO\textsubscript{4} to understand the effect of Mn-site substitution on the process of  Na migration through this material.
\end{abstract}

\section{Introduction}
The last two decades witnessed significant advancements in the field of Li-ion battery as the properties like energy density, rate capacity and cost of the material had considerable improvements. But the ever increasing demand and geographically restricted Li resources impose constraints to the cost-effective production of  Li-ion battery\cite{turcheniuk2018ten,xu2020future}. Thus materials scientists were encouraged to think about non-Li-based batteries, and Na-ion battery was an apparent alternative candidate as it has been developing along with the Li-ion battery since its inception in the 1990s. Furthermore, the need for the non-Li-based battery was inevitable as the field of clean and environmentally friendly energy production was blooming over the years, and it requires safe and economically feasible energy storage mechanisms. The sole candidate, Li-ion battery, may not be able to satisfy this increasing need. Even though advancements in Na-ion technology could improve the situation, the intrinsic downside of Na-ion battery technology is its low energy density compared to Li-ion battery. The Na ions are three times heavier than Li ions and the redox potential of Na (–2.71 V vs SHE) is 0.3 V lower than that of Li (–3.02 V vs SHE). Thus, the development of the Na-ion battery technology is mainly focused on large-scale stationary applications such as grid energy storage where the energy density or weight of the battery system is not a necessary limiting factor\cite{tarascon2020ion}. Clearly, the Na-ion battery is not expected to completely replace the Li-ion battery but to complement it to meet the ever-increasing demand for energy storage materials.
\par Due to the growing societal demands on sustainability, the past decade witnessed intensifying research on developing high-efficiency Na-ion batteries, considering the availability and abundance of Na resources across the world. Also, aluminum can be used as a current collector in Na-ion batteries, but, not in Li-ion batteries as it forms alloys with its cathode materials. Copper is used instead, which is costlier than aluminum\cite{hwang2017sodium}. Moreover, it has to be noted that the Na-ion battery chemistry is equivalent to that of Li-ion, and thus the scientific knowledge acquired over the past years for Li-ion battery can be easily transferred to Na-ion battery development. The research on cathode materials for Na-ion battery was mainly focused on layered materials until the introduction of polyanionic compounds which showed promising characteristics due to their stability, safety, and suitable operating voltages \cite{ni2017polyanion, fang2015high, saravanan2013first}. Among the wide variety of polyanionic compounds considered as cathode materials for Na-ion batteries, the Mn containing silicate such as Na\textsubscript{2}MnSiO\textsubscript{4} showed higher capacities with an added benefit of utilization of more than one Na atom per formula unit in the redox process. Furthermore, Si is the second most abundant element in the earth's crust\cite{RUDNICK20141} and various silicate based polyanions can be readily synthesized by relatively simple methods\cite{panigrahi2017sodium}. Another advantage is the usage of environmentally friendly and cost effective Mn as the transition metal. 
Studies on the Li counterpart of Na\textsubscript{2}MnSiO\textsubscript{4}  demonstrated that the removal of more than one Li leads to structural degradation and the electrochemical studies revealed that the higher capacities up to 270 mAhg\textsuperscript{-1} were reached only sporadically\cite{kokalj2007beyond}.
\par 
Na\textsubscript{2}MnSiO\textsubscript{4} has been reported to be used for the synthesize of metastable \textit{Pn}-Li\textsubscript{2}MnSiO\textsubscript{4} by ion-exchange method\cite{duncan2011novel}. The  first electrochemical study on Na\textsubscript{2}MnSiO\textsubscript{4} as a cathode material for Na-ion battery was done by Chih-Yao Chen \textit{et al.} and obtained a reversible capacity of around 125 mAhg\textsuperscript{-1}\cite{chen2014na2mnsio4} based on the removal of 0.9 Na from formula unit of the material. The synthesis of Na\textsubscript{2}MnSiO\textsubscript{4} from high-temperature solid-state reaction method was done by Xia \textit{et al.} and they have reported that it can be used as Na-ion capacitor based on its electrochemical study\cite{xia2017electrochemical}. A sol-gel synthesized Na\textsubscript{2}MnSiO\textsubscript{4}/C/G composite was used as a cathode material to deliver a capacity as high as 180 mAhg\textsuperscript{-1} and also it exhibited high rate performance\cite{zhu2017facile}. The improved properties were attributed to the inclusion of graphene. 
\par Subsequent electrochemical storage studies on Na\textsubscript{2}MnSiO\textsubscript{4} registered  highest ever reported specific capacity of 210 mAhg\textsuperscript{-1} for a polyanionic compound with the utilization of 1.5 electrons per formula unit\cite{law2017na2mnsio4}. The increase in capacity is attributed to the additive vinylene carbonate, which stabilizes the electrode-electrolyte interface during cycling. The computational studies on this compound are mainly focussed on the diffusion analysis of Na through the host structure\cite{kuganathan2018defects, zhang2015ion}. Zhang \textit{et al.}\cite{zhang2015ion} studied the diffusion of both Li and Na in  Na\textsubscript{2}MnSiO\textsubscript{4} in order to investigate the functioning of the material as a hybrid ion battery cathode material\cite{barker2006hybrid}. 
\par In this work, we present the thermodynamic and electronic structure study of Na\textsubscript{2}MnSiO\textsubscript{4} using density functional theory (DFT) based calculations. Cluster expansion assisted density functional theory is used to obtain the $0-K$ voltage profile and Monte Carlo simulation to obtain the finite temperature voltage profile. This combined approach  is proven to be effective in determining the thermodynamics of several cathode materials for Li-ion and Na-ion batteries\cite{kaufman2019x, toriyama2019potassium, kolli2018first, chang2016li}. The investigation of the voltage profile and the related characteristics using cluster expansion is not applied for this compound so far. For further understanding of the electronic structure and the redox reactions involved in the electrochemical activity, we have performed the electronic structure calculations for Na\textsubscript{2}MnSiO\textsubscript{4} and its intermediate ground states configurations during the redox reaction. Charge density, charge transfer, electron localization function (ELF) and the Bader charges are analyzed to have a clear understanding of the bonding properties and the amount of charge associated with each species during the battery operation. In view of the involvement of oxygen in the redox process, the possibilities of oxygen evolution is also probed for various charging stages. In addition to such studies, we have also performed the Na diffusion analysis in pure and the partial substitution of Mn by Fe in Na\textsubscript{2}MnSiO\textsubscript{4} to identify the involvement of lattice and electronic structure change on the Na-ion migration.

\section{Computational Details}
All calculations have been performed within the frame work of DFT using the software Vienna Ab-initio Simulation Package(VASP)\cite{kresse1996efficiency, kresse1996efficient, kresse1993ab}. The projected augmented wave method\cite{blochl1994projector, kresse1999ultrasoft} implemented in VASP is used to describe the interaction between valence and core electrons. Plane-wave basis sets were used to describe the valence electrons up to the cut off energy of 520\,eV. $\Gamma$-centered grid of \textbf{k} point sampling has been done for all the calculations. Graphics Processing Unit (GPU) port of VASP is used to do all the calculations except the diffusion barrier calculations\cite{hacene2012accelerating, hutchinson2012vasp}. Spin-polarized calculations were made with the assumption of ferromagnetic ordering in the transition metal sub-lattice. To account for correlation effect of the localized \textit{d} orbitals of the transition metals, Hubbard $U$ value of 3.9\,eV for Mn and 2.9\,eV for Fe is included. Generalized gradient approximation (GGA) functional formulated by Perdew-Burkhe and Ernzerhoff (PBE)\cite{perdew1996generalized} was used to account for the exchange and correlation energy. The ionic positions and lattice parameters of each structure were fully relaxed. 
\par We have used the cluster expansion Hamiltonian to find the total energies of different Na-vacancy configurations for varying concentrations of Na in Na\textsubscript{2}MnSiO\textsubscript{4} host. The cluster expansion parametrizes the formation energy as a polynomial in terms of the occupational variables.
\begin{equation} \label{cl_eqn1}
E(\sigma) = V_0+\sum_{\alpha}V_{\alpha}\phi_{\alpha}
\end{equation}
Where the cluster functions 
\begin{equation} \label{cl_eqn2}
\phi_\alpha(\sigma) = \prod_{i \in \alpha} \sigma_i \
\end{equation}
are the products of occupation variables $\sigma$ corresponding to each cluster of Na sites(point cluster, pair cluster, triplet cluster, etc.). The occupation variables are assigned $+$1 or $-$1 depending on whether a site is occupied by Na or not. Even though the number of clusters possible in a system is huge, the cluster expansion usually truncates with the inclusion of a maximal  number of clusters. In this study, a training set of 190 configurations equipped with the  DFT calculated total energies have been used to parametrize the cluster expansion and to obtain the effective cluster interactions (ECI) which are the expansion coefficients $V_0$, $V_\alpha$ in the cluster expansion Hamiltonian shown in equation \ref{cl_eqn1}. The resultant cluster expansion contains 63 different ECI values with the rms error between the calculated and cluster expanded formation energies of 23.8 meV/f.u and with a cross validation (CV) score of 32 meV/f.u. The CV score is a measure of the predictability of the cluster expansion. This quantifies the maximum error in predicting the formation energy of a new configuration not included in the training set. i.e., the CV score is a measure of how well the cluster expansion can predict the formation energy of a completely new configuration.  The obtained cluster expansion Hamiltonian is used in the grand canonical Monte Carlo simulation to predict the finite temperature properties and to obtain voltage profile at room temperature. The Clusters Approach to Statistical Mechanics (CASM)\cite{thomas2013finite, van2010linking, puchala2013thermodynamics} software package is used to construct and parametrize the cluster expansion and to do Monte Carlo simulations. 
\par The lowest energy configurations for different concentrations of Na in Na\textsubscript{2x}MnSiO\textsubscript{4} were considered for further analysis, such as electronic structure, Bader effective charge, \cite{sanville2007improved, henkelman2006fast} and Crystal Orbital Hamilton population (COHP)\cite{deringer2011crystal, dronskowski1993crystal, nelson2020lobster} analyses. Nudged elastic band (NEB) method\cite{jonsson1998nudged} is used to find the diffusion barriers for Na migration in pure and the partially Fe substituted Na\textsubscript{2}MnSiO\textsubscript{4}. A supercell of 2$\times$2$\times$2 dimension is used for all the NEB calculations to ensure sufficient spacing between vacancies. Converged NEB calculations are restarted with Climbing Image NEB (CI-NEB)\cite{henkelman2000climbing, sheppard2008optimization} settings to improve the accuracy of barrier height. For computational simplicity all the calculations are done in the ferromagnetic configuration though the system prefers to have an anti-ferromagnetic ground state. However, the energy difference between the ferromagnetic and anti-ferromagnetic systems is less than 7 meV/fu and hence assuming ferromagnetic ordering will not affect our calculated results significantly.  Moreover, the thermal excitations, which are of the order of magnitude of k\textsubscript{B}T= 25 meV at room temperature, would break the magnetic ordering in the system.

\section{Results and Discussion}
\subsection{Structural details}
\begin{figure}[h]
\centering
\includegraphics[scale=0.75]{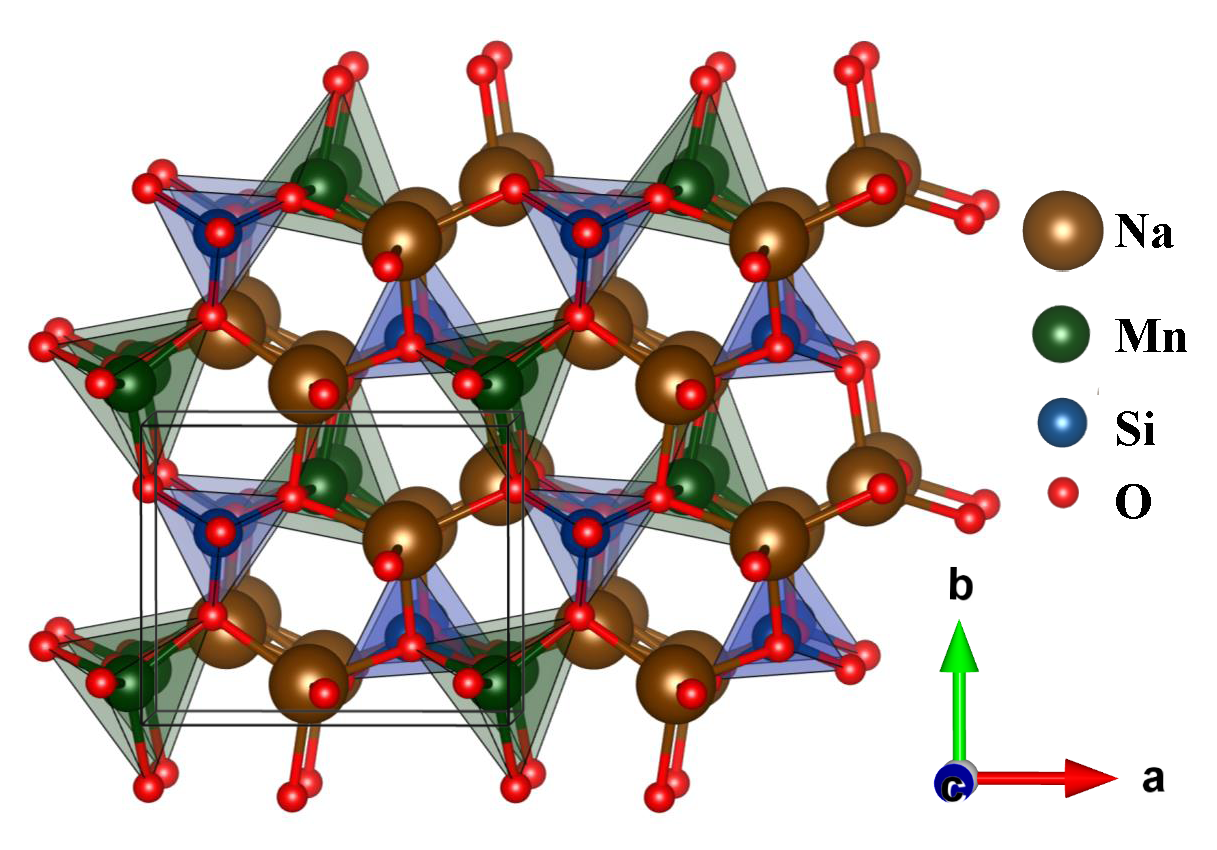}
\caption{Optimized crystal structure of Na\textsubscript{2}MnSiO\textsubscript{4} displaying the interconnected tetrahedral network. MnO\textsubscript{4} and SiO\textsubscript{4} tetrahedra are shown in light green and light blue colours, respectively. This picture has been created using the VESTA software\cite{momma2011vesta}}
\label{fig:1}
\end{figure}

The experimental XRD measurements show that Na\textsubscript{2}MnSiO\textsubscript{4}  stabilize in a  monoclinic structure with the  space group \textit{P1n1}\cite{nalbandyan2017a2mnxo4}. This structure has  3D connected corner shared MnO\textsubscript{4} and SiO\textsubscript{4} tetrahedra with Na atoms in the voids as shown in Fig. \ref{fig:1}. All three cations (Na, Si, Mn) in this material are tetrahedrally connected with oxygen atoms and forming a double layered hexagonal sub-lattice. The cations are in a distorted tetrahedral environment and all the tetrahedra are pointing in the same direction. Each Na tetrahedron in this structure corner share its oxygen with neighbouring  Na, Si, and Mn tetrahedra. In other words, every oxygen atom is shared between two NaO\textsubscript{4}, one MnO\textsubscript{4}, and one SiO\textsubscript{4} tetrahedra.  The optimized structural parameters using force as well as stress minimization are given in the Table \ref{table:1} along with the corresponding experimental structural parameters for comparison. Certain lattice parameters are overestimated by up to 1.6 \% possibly due to the application of GGA functional. However, from  Table \ref{table:1} one can find a good agreement between the calculated and experimental lattice parameters.

\begin{table}[h]
\small
  \caption{The calculated and experimental equilibrium lattice parameters for  Na\textsubscript{2}MnSiO\textsubscript{4} in the  monoclinic structure with the  space group \textit{P1n1}.}
  \label{table:1}
  \begin{tabular}{llll}
    \hline
    Parameter & Calculated & Experimental\cite{nalbandyan2017a2mnxo4} & Percentage error\\
   \hline

     a (in \AA) & 6.99 & 7.03 & 0.57 \\
     b (in \AA) & 5.70 & 5.61 & 1.6 \\
     c (in \AA)& 5.39 & 5.34 & 0.94 \\
     $\beta$ & 89.75 & 89.79 & 0.04 \\

   \hline
  \end{tabular}
\end{table}

\subsection{Calculation of formation energy}
The cluster expansion method can be used to find any material property that depends on the configurational arrangement of atoms in the lattice, such as the free energy, entropy, formation energy and many other thermodynamic quantities\cite{asta1991effective, wolverton1991effective, sanchez2010cluster}. Here, the formation energy is considered as the configuration dependent quantity and is expanded in terms of cluster basis functions multiplied by expansion coefficients that define whether a particular site is occupied by Na or not. The formation energy not only gives an effective way to understand the relative stability of different phases but also gives insight into the electrochemical properties of the material. The cell reaction for the compound Na\textsubscript{2}MnSiO\textsubscript{4} with two Na atoms per formula unit is 
\begin{equation} \label{eu_eqn3}
Na\textsubscript{2x\textsubscript{1}}MnSiO\textsubscript{4} \longrightarrow Na\textsubscript{2x\textsubscript{2}}MnSiO\textsubscript{4} + 2(x\textsubscript{1} -x\textsubscript{2}) Na
\end{equation}

\begin{figure}[h]
    \includegraphics[scale=1]{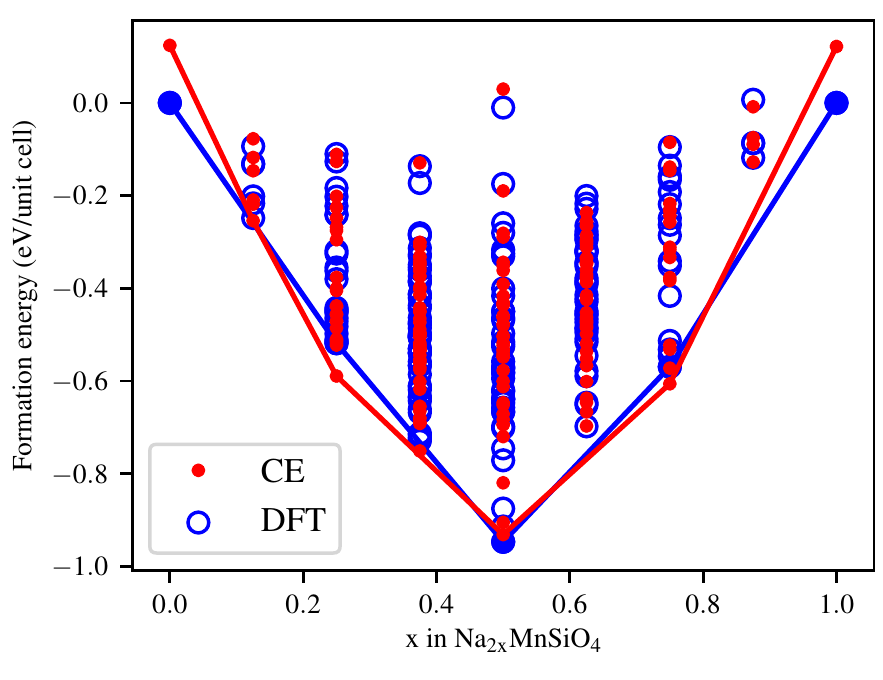}
    \caption{The formation energies per unit cell of Na\textsubscript{2x}MnSiO\textsubscript{4} is calculated for 190 configurations by DFT are shown in blue open circles and the corresponding formation energies found from cluster expansion (CE) method are shown as red dots. The ground state configurations are connected by a red line for cluster expanded formation energies and the ground state configurations from DFT calculations represented by blue dots connected by a blue line.}
    \label{fig:2}
\end{figure}

The formation energy of any arbitrary concentration $x$ of Na in Na\textsubscript{2x}MnSiO\textsubscript{4} can be written as 
\begin{equation}\label{equn_4}
   E\textsubscript{f}(Na\textsubscript{2x}MnSiO\textsubscript{4})=  E(Na\textsubscript{2x}MnSiO\textsubscript{4}) - x E(Na\textsubscript{2}MnSiO\textsubscript{4})
- (1 -x) E(MnSiO\textsubscript{4}) 
\end{equation} 
\label{eu_eqn1}
Where $E\textsubscript{f}$ is the formation energy of the compound with $2x$ Na concentration. E(Na\textsubscript{2x}MnSiO\textsubscript{4}) is the total energy of the material with an intermediate $2x$ Na concentration. $E(Na\textsubscript{2}MnSiO\textsubscript{4})$ and $E(MnSiO\textsubscript{4})$ are the total energies of sodiated and desodiated materials, respectively. Different Na-vacancy arrangements for each concentration of Na in the host material up to a supercell size of 2 units were constructed using the CASM code. The formation energy for 190 different configurations is calculated using VASP and is used to construct the cluster expansion. The cluster expanded (CE) and DFT-based formation energies are shown in Fig. \ref{fig:2}. The formation energies found from both the methods agree fairly well with each other for most of the configurations. The CV score for the cluster expansion is 4 meV~atom\textsuperscript{-1} and the rms error is below 3 meV~atom\textsuperscript{-1}. Thus the constructed cluster expanded Hamiltonian is reasonably accurate to consider for further calculations to predict the formation energy of new compositions within the range.

\par The tie line that connects the common tangents of lowest-lying configurations forms a hull-like structure and is generally referred to as the formation energy hull\cite{ishikawa2020evolutionary}. The slope of the formation energy hull is a direct indication of the amount of average voltage of the cathode material. The convex hull is symmetrically shaped between the dilute and non-dilute Na concentrations. Negative formation energies indicate that the particular configuration is stable with respect to the two phase mixture of the sodiated and desodiated configurations. Numerous closely lying configurations near the ground state configurations for all Na concentrations indicate that the corresponding voltage curve may have solid solution behaviour at higher temperatures. This can be observed in the finite temperature voltage-profile curve in the next section.
\subsection{Estimation of voltage profile}
\begin{figure}[h]
    \includegraphics[scale=0.5]{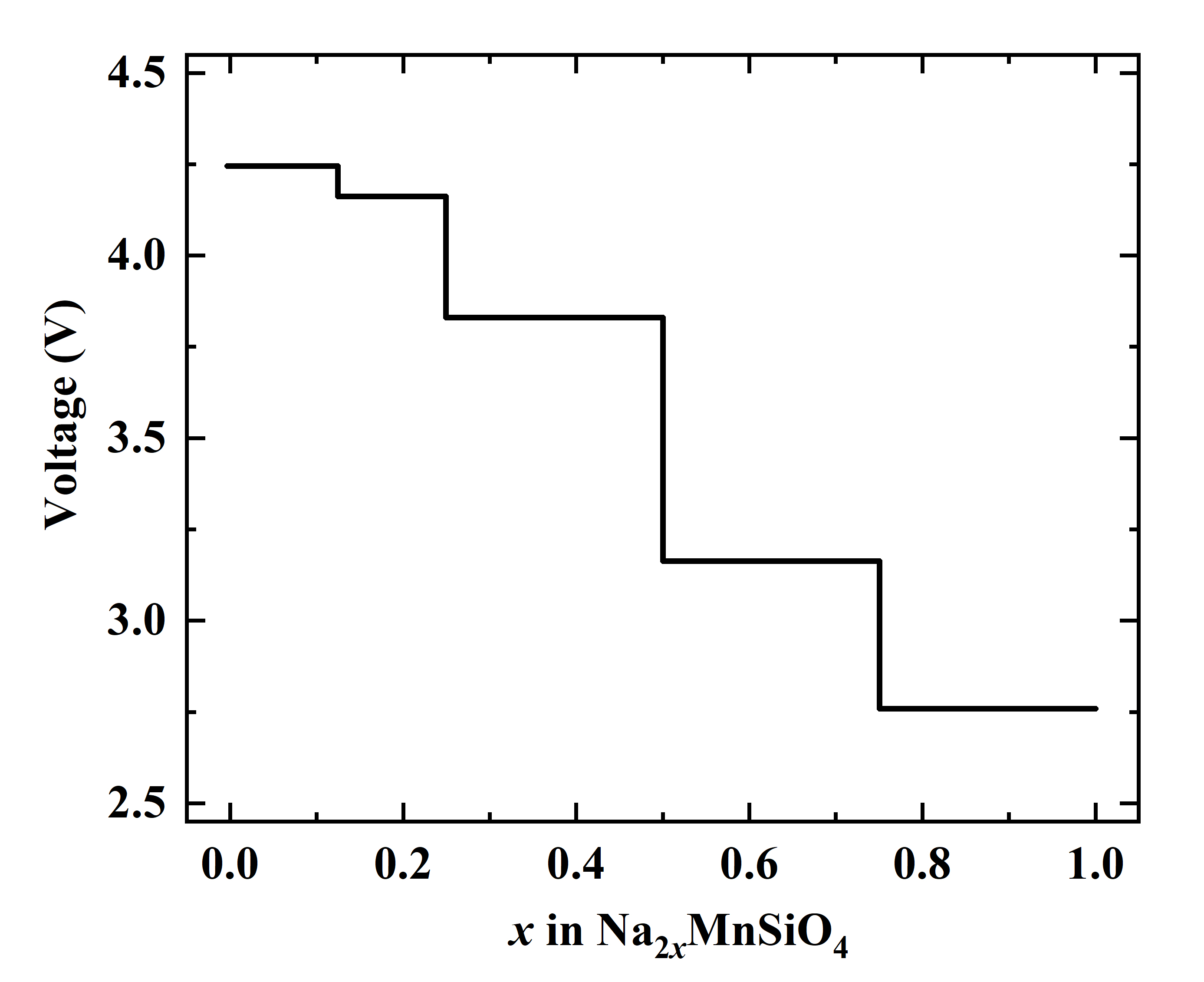}
    \caption{The voltage profile at $0 K$ calculated directly from the output of convex hull obtained by the cluster expansion using equation \ref{eu_eqn5}.}
    \label{fig:3}
\end{figure}

\begin{figure}
    \includegraphics[scale=0.5]{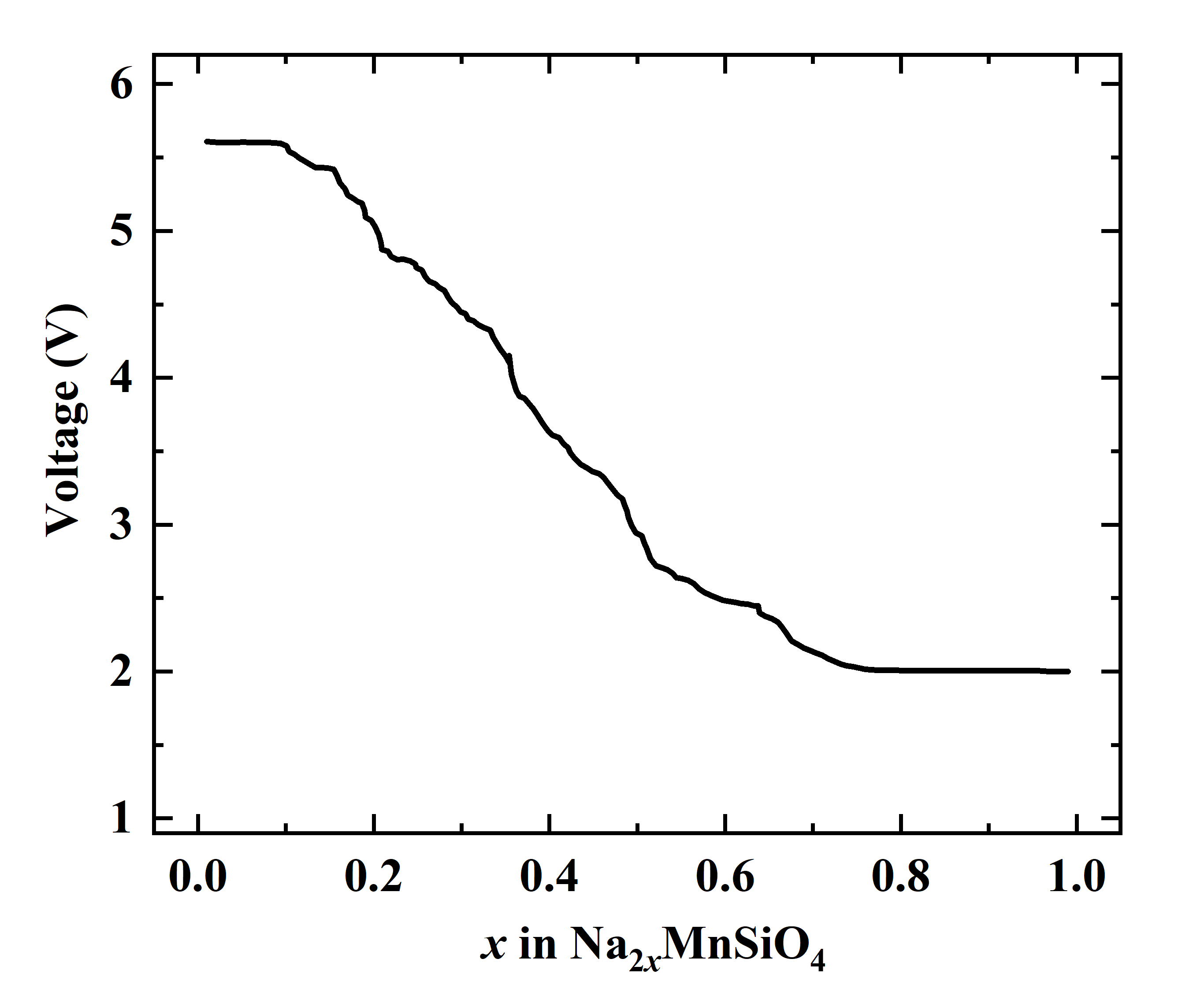}
    \caption{The voltage profile at $300~K$ calculated directly from the output of Monte Carlo simulation using equation \ref{eu_eqn6}.}
    \label{fig:4}
\end{figure}

The compound Na\textsubscript{2}MnSiO\textsubscript{4} contains 2 Na atoms per formula unit and hence, the voltage equation corresponding to the cell reaction as in equation \ref{eu_eqn3} can be written as
\begin{equation}\label{eu_eqn5}
    V = -\frac{E(Na\textsubscript{2x\textsubscript{1}}MnSiO\textsubscript{4})- E(Na\textsubscript{2x\textsubscript{2}}MnSiO\textsubscript{4})- 2(x\textsubscript{1} -x\textsubscript{2}) E(Na)}{2(x\textsubscript{1} -x\textsubscript{2})}
\end{equation}
The voltage profile at $0~K$ from the formation energy curve is obtained from the above relation. Once the intermediate ground states are identified, the calculation of the voltage between two consecutive Na concentrations is done with the use of total energies of the corresponding ground states.  The voltage profile of the cathode material Na\textsubscript{2}MnSiO\textsubscript{4} is obtained from the formation energy curve using the equation \ref{eu_eqn5} as shown in Fig. \ref{fig:3}, as well as from the direct output of the Monte Carlo simulations in Fig.\ref{fig:4}. The output of the Monte Carlo simulation gives the chemical potential of the intercalating ion as a function of the Na concentration. The voltage of Na ion battery is related to the chemical potential difference of Na in anode and cathode as defined in the equation
\begin{equation} \label{eu_eqn6}
V  = -\frac{\mu\textsubscript{cathode} - \mu\textsubscript{anode}}{zF}
\end{equation}
The chemical potential  of Na in the anode ($\mu \textsubscript{anode}$) is approximated to be the total energy of \textit{bcc} Na. The chemical potential of Na in the cathode material ($\mu\textsubscript{cathode}$) can be obtained directly from the  Monte Carlo simulations, and $z$ is the number of electrons transported by each intercalating ion. F is Faraday's constant.
Using the above approach the estimated voltage profile at $300~K$ is shown in Fig. \ref{fig:4}.
Though many factors such as the kinetic effects of the intercalating ion and chemical nature of the electrolyte can affect the voltage curve; we present the voltage curve as obtained from the formation energy convex hull as a qualitative measure for the study of thermodynamic properties of this cathode material. All sources of dissipation are neglected in calculating the voltage curve displayed in Fig. \ref{fig:4}, and ideal electrolyte, as well as counter electrodes, are assumed. The significant feature of this curve is the observance of a solid solution region spanning between $x$ = 0.1 to $x$ = 0.75 in Na\textsubscript{2}MnSiO\textsubscript{4} and pronounced miscibility gap at the extreme concentrations. The solid solution region contains many small steps indicating the possibility of ordering between the Na and vacancy at these concentrations. Since the steps are fairly small, the ordering are stable only in the minute window of chemical potential range. The plateau-like region at concentrations above 0.75 Na is an indication of two phase region and the consequent occurrence of a miscibility gap. The plateau region in the voltage profile is a distinctive feature that can also be observed in the experimental first charging curve of Na\textsubscript{2}MnSiO\textsubscript{4} at various operating temperatures\cite{chen2014na2mnsio4}.

\par The observable difference between the $0~K$ voltage profile and the $300~K$ voltage profile is primarily due to the introduction of more intermediate ground states for finite temperature calculations that are not obtainable for the initial $0~K$ calculation due to the limited cell size considered.

\subsection{Electronic structure analysis}
\begin{figure}[h]
\centering
\includegraphics[scale=0.5]{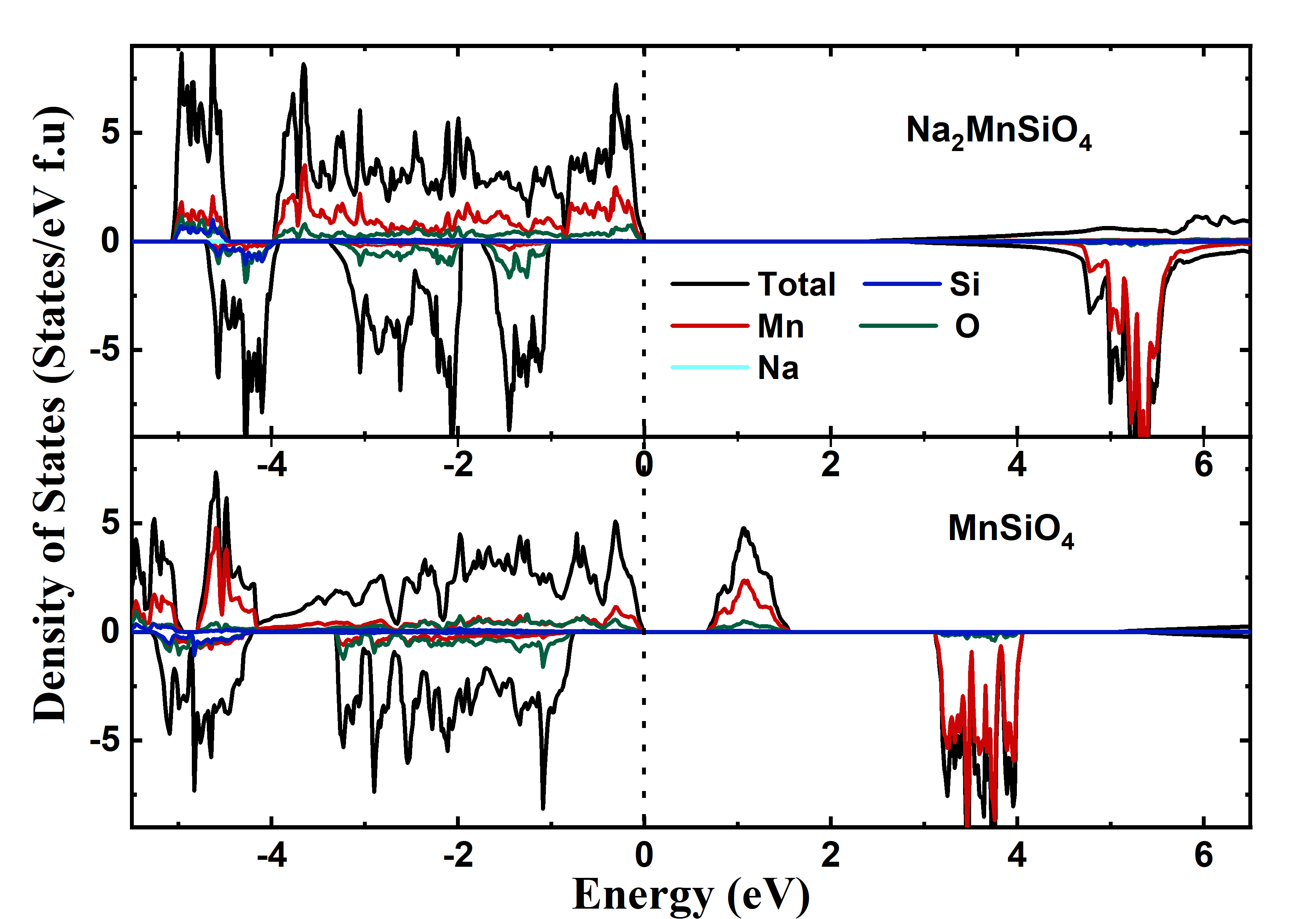}
\caption{Total and atom projected density of states of Na\textsubscript{2}MnSiO\textsubscript{4} and MnSiO\textsubscript{4}}
\label{fig:5}
\end{figure}
\begin{figure}[h]

\includegraphics[scale = 0.5]{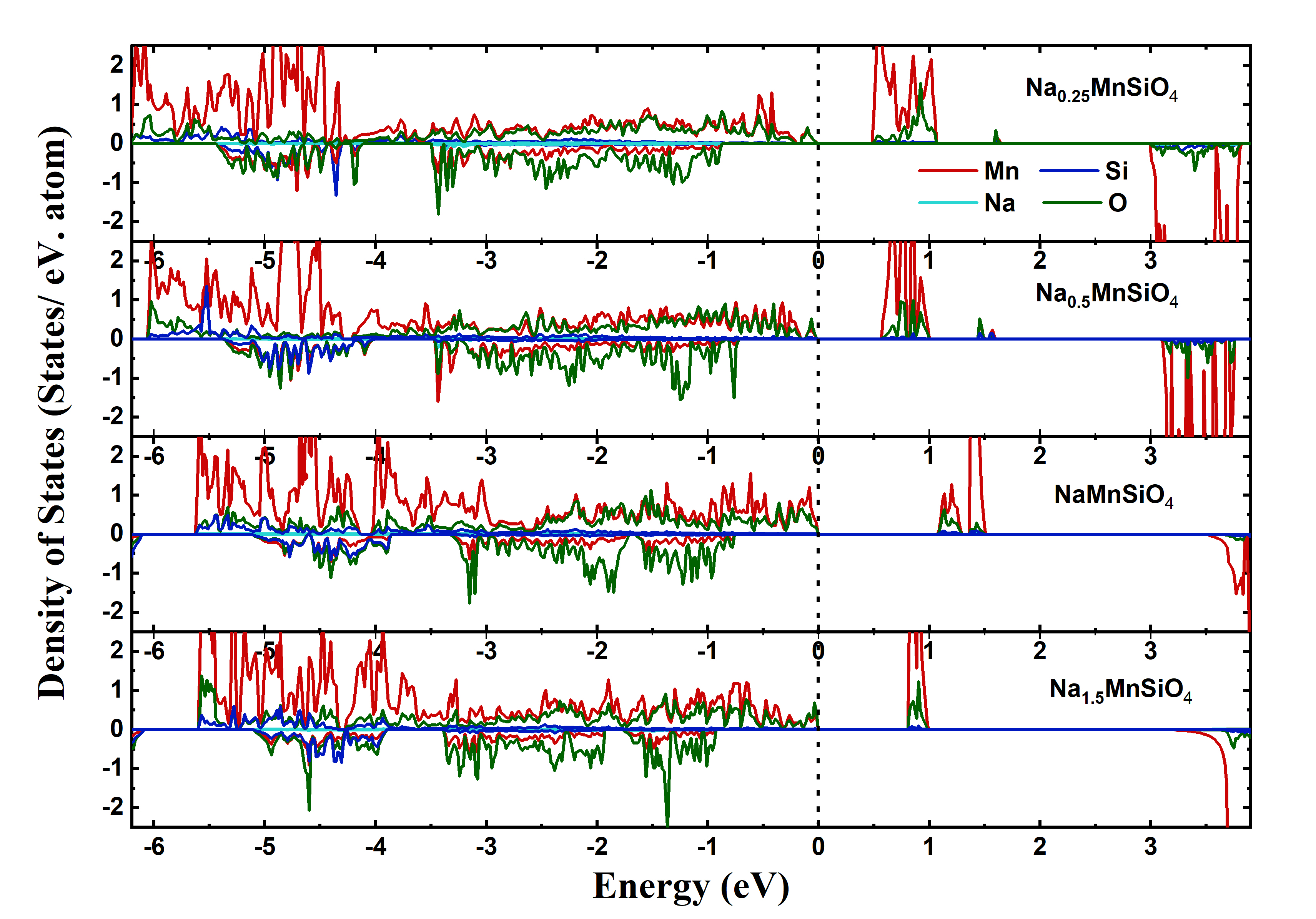}
\centering
\caption{Atom projected density of states of intermediate compounds of Na\textsubscript{2}MnSiO\textsubscript{4}}
\label{fig:6}
\end{figure}

\par We have plotted the total as well as projected density of states (DOS) for Na\textsubscript{2}MnSiO\textsubscript{4} and the completely desodiated compound MnSiO\textsubscript{4} as shown in Fig: \ref{fig:5}. The significant difference in the electronic structure between the sodiated and desodiated system is that in the desodiated system, an intermediate band (IB), made up of O-\textit{p} and Mn-\textit{d} states are formed, which reduces the bandgap drastically. The bandgap measured in the DOS plot of Na\textsubscript{2}MnSiO\textsubscript{4} is 2.4~eV and that of the desodiated compound is 0.7~eV. We can assume a decrease in band gap and an increase in electrical conductivity as the desodiation progresses. The major contribution to the spin-up valence band maximum (VBM) is from the Mn-\textit{d}-states while that for the spin-down VBM is from the O-\textit{p}-states in the fully sodiated system. However, as shown in Fig: \ref{fig:5}, the spin-up VBM for completely desodiated system has Mn-\textit{d}-states with substantial O-\textit{p}-states while the spin-down VBM has major contributions from O-\textit{p}-states similar to the fully sodiated phase. The contributions to the conduction band edge of  Na\textsubscript{2}MnSiO\textsubscript{4} is  from  the spin-down states of Mn$-$\textit{d} electron. In contrast, the conduction band edge of the desodiated base is contributed by Mn up-spin \textit{d} electrons with a small contribution from O$-$\textit{p} states. 

\par The valence band edge of the compound Na\textsubscript{2}MnSiO\textsubscript{4} is dominated by Mn up-spin electrons, and the down spin electrons are around 1~eV below the VBM, indicating that the majority spin electrons are mainly contributing to the redox reaction. The unoccupied states at the band edge were well  dispersed over a large energy range for Na\textsubscript{2}MnSiO\textsubscript{4}. However, in the completely desodiated phase, these unoccupied states in the band edge are in a narrow energy range between 0.7~eV and 1.6~eV in the spin-up states. For both Na\textsubscript{2}MnSiO\textsubscript{4} and MnSiO\textsubscript{4}, within the valence band, the spin-up states of both Mn and O atoms fall in the same energy range, and hence they are expected to form a directional covalent bond. On the other hand, due to magnetic polarization, the Mn$-$\textit{d} states are moved to the conduction band in the minority spin state. However, due to nonmagnetic nature of O, its \textit{p}-states are distributed in the minority spin channel of the valence band also.

\par The energetically degenerate electronic states of Si and O between $-$4~eV and $-$5~eV for Na\textsubscript{2}MnSiO\textsubscript{4} and between $-$4~eV and $-$6~eV for MnSiO\textsubscript{4} implies the possibility of forming covalent interaction between them. Also, these states are present deep inside the valence band and thus will not take part in the chemical reaction.
\par The atom projected DOS for the intermediate ground states with the composition $x$ = 0.125, 0.25, 0.5, 0.75 in  
 Na\textsubscript{2x}MnSiO\textsubscript{4} for their ground states obtained from the cluster expansion method described in the previous section, are shown in Fig. \ref{fig:6}. All the intermediate compounds follow the same characteristics as that of MnSiO\textsubscript{4} with the formation of the intermediate band states originating from spin-up electronic states of Mn$-$\textit{d} states with noticeable contribution from O$-$\textit{p}-states. This implies that the desodiation process drastically changes  the optical properties of this material. The contribution from O to the IB states progress as the desodiation process continues from Na\textsubscript{0.5}MnSiO\textsubscript{4} to Na\textsubscript{0.25}MnSiO\textsubscript{4}, as evident from Fig. \ref{fig:6}.  Furthermore, if the Na concentration decreases, the IB width systematically increases. So we expect that by measuring the lower energy optical excitation spectra during the redox reaction, one can measure the degree of desodiation in this system. 
 
 \par Both sodiated and desodiated phases, the Mn atoms are surrounded by O atoms tetrahedrally and also Mn$-$\textit{d} states and the O$-$\textit{p} states are energetically degenerate in the entire valence band indicating the presence of covalent bonding between them. However, the peaks in the Mn$-$\textit{d} states in the VB are well localized around  $-$5.5~eV to $-$3.5~eV, and the O$-$\textit{p} states are dominantly present in the higher energy region of the valence band indicating the presence of noticeable ionic bond, which is established by other bonding analysis discussed later. So the bonding interaction between the Mn and O is of mixed iono-covalent in nature. The Si atoms are tetrahedrally coordinated by O atoms and also Si$-$\textit{p} and O$-$\textit{p} states are energetically degenerate in the entire VB for both spin channel with substantial contribution around $-$5.5~eV to $-$3.5~eV energy range. 

\begin{figure}[h]
\includegraphics[scale = 0.6]{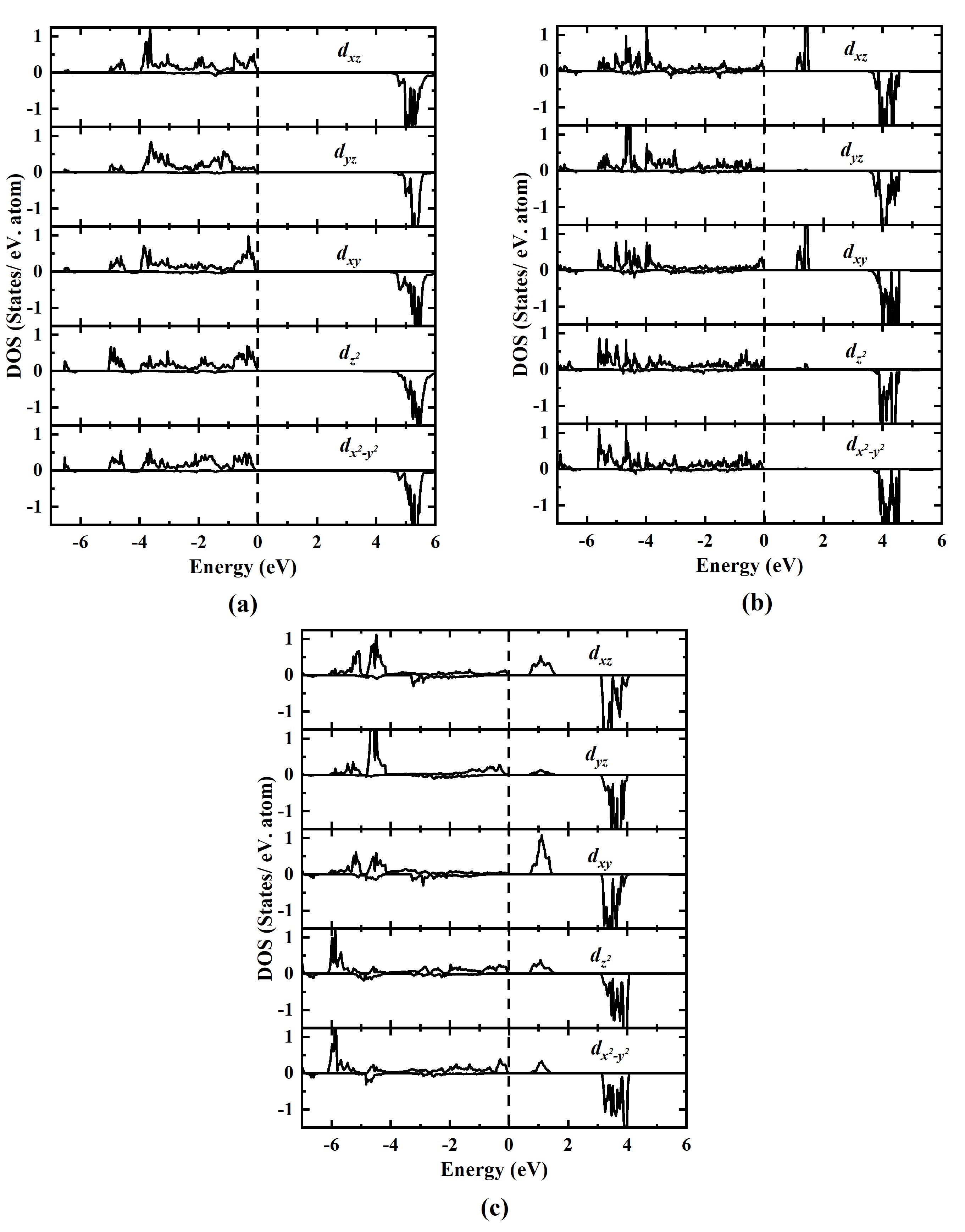}
\centering
\caption{Orbital decomposed DOS of the Mn$-$\textit{d} states for Na\textsubscript{2}MnSiO\textsubscript{4}, NaMnSiO\textsubscript{4} and MnSiO\textsubscript{4}}
\label{fig:7}
\end{figure}
\par In order to understand the change in the oxidation state of Mn ions as a function of desodiation process in Na\textsubscript{2}MnSiO\textsubscript{4}, we have plotted the  orbital decomposed Mn$-$\textit{d} DOS for  Na\textsubscript{2}MnSiO\textsubscript{4}, NaMnSiO\textsubscript{4} and MnSiO\textsubscript{4} in Fig \ref{fig:7}. From this figure, it is clear that in the case of Na\textsubscript{2}MnSiO\textsubscript{4}, the Mn ions are in the high spin state and hence all the five \textit{d} orbitals are half-filled, resulting in the net magnetic moment of $4.6~\mu_B$/Mn. Also, these results show that the Mn ion is in the $+2$ oxidation state. When we desodiate the system to NaMnSiO\textsubscript{4}, the orbital projected Mn$-$\textit{d} states show that all the five \textit{d} orbitals are almost equally filled, and also, the \textit{d} states get localized in the lower energy part of the valence band. Further, due to the transfer of electrons from Mn to the O during the desodiation process, additional empty states are formed in the Mn$-$\textit{d\textsubscript{xy}} and \textit{d\textsubscript{xz}} orbitals as evident from Fig. \ref{fig:7}. If we altogether remove the Na from the system and attain MnSiO\textsubscript{4}, all  the five Mn$-$\textit{d} states are further localized, and minimal Mn$-$\textit{d} states are present in the top of the valence band. Moreover, due to additional transfer of electrons from Mn to O during desodiation, additional empty \textit{d} states are formed in the \textit{d\textsubscript{x\textsuperscript{2}-y\textsuperscript{2}}} and \textit{d\textsubscript{z\textsuperscript{2}}} orbitals. Due to the transfer of electrons as well as the filling of the minority spin channel in the valence band, the magnetic moment at the Mn site is reduced from  $4.6~\mu_B$/Mn in Na\textsubscript{2}MnSiO\textsubscript{4} to $3.2~\mu_B$/Mn  in MnSiO\textsubscript{4}. 

\par The  tetrahedral crystal field from O atoms around  Mn  lift the degeneracy of the \textit{d} states. The tetrahedral field splits the \textit{d} states in to energetically lower e\textsubscript{\textit{g}}  (\textit{d\textsubscript{x\textsuperscript{2}-y\textsuperscript{2}}}, \textit{d\textsubscript{z\textsuperscript{2}}}) and higher t\textsubscript{\textit{2g}} (\textit{d\textsubscript{xy}}, \textit{d\textsubscript{xz}}, \textit{d\textsubscript{yz}}) states. For all three compounds, the orbital decomposition of Mn$-$\textit{d} states reveals that the filling up of the electrons in the valence band region for both \textit{d\textsubscript{x\textsuperscript{2}-y\textsuperscript{2}}} and \textit{d\textsubscript{z\textsuperscript{2}}} orbitals are almost the same, indicating their degeneracy and the other three orbitals are also occupied in a similar manner. This substantiates the presence of the above mentioned tetrahedral field splitting throughout the desodiation process. One can also observe that the occupancy of e\textsubscript{\textit{g}} states are towards the lower energy part of valence band region in the energy spectrum and the  t\textsubscript{\textit{2g}} orbitals are filled more towards the higher energy in the valence band region indicating the existence of the slight energy difference between the e\textsubscript{\textit{g}} and  t\textsubscript{\textit{2g}} states in the tetrahedral field. Nevertheless, the splitting is not as prominent as one would expect because of the presence of covalent interaction between the Mn and O in all these phases. Due to the covalency mentioned above, the Mn \textit{3d} states are distorted heavily and make the crystal field splitting less visible. It has to be noted that, even though the tetrahedral environment is distorted by the charge transfer during the desodiation process, the Mn atoms remain in their tetrahedral coordination despite the fact that, Mn\textsuperscript{4+} ions prefer an octahedral environment\cite{martins2017local, beji2015polyol, chen1996size}.

\par In Na\textsubscript{2}MnSiO\textsubscript{4}, the Mn\textsuperscript{2+} ions  are in high spin state with all of their spin-up states are occupied and the spin-down states are unoccupied, as seen from the orbital decomposed DOS plot. As the Na ions are extracted out of the compound, the Mn ions undergo changes in their oxidation state to adjust the deficiency of electronic charge density from the outgoing electrons. This creates vacant spin-up states in the conduction band, as can be observed as the intermediate \textit{d} states in the fig \ref{fig:7}. Moreover, the \textit{p} states of O also contribute to the IB formed while the discharging process happens. This suggests that the O atoms are also involved in the redox process in this system as evident in the site projected DOS depicted in fig \ref{fig:6}. Due to the formation of IB states, the bandgap  gets reduced and as a consequence of that, optical absorption is expected to start even at very low energy. So our study suggests that by optical measurements also one can monitor the desodiation process.

 \subsection{Charge density, Charge transfer, and ELF analyses}           

\par
\begin{figure*}[h]
\includegraphics[scale = 0.9]{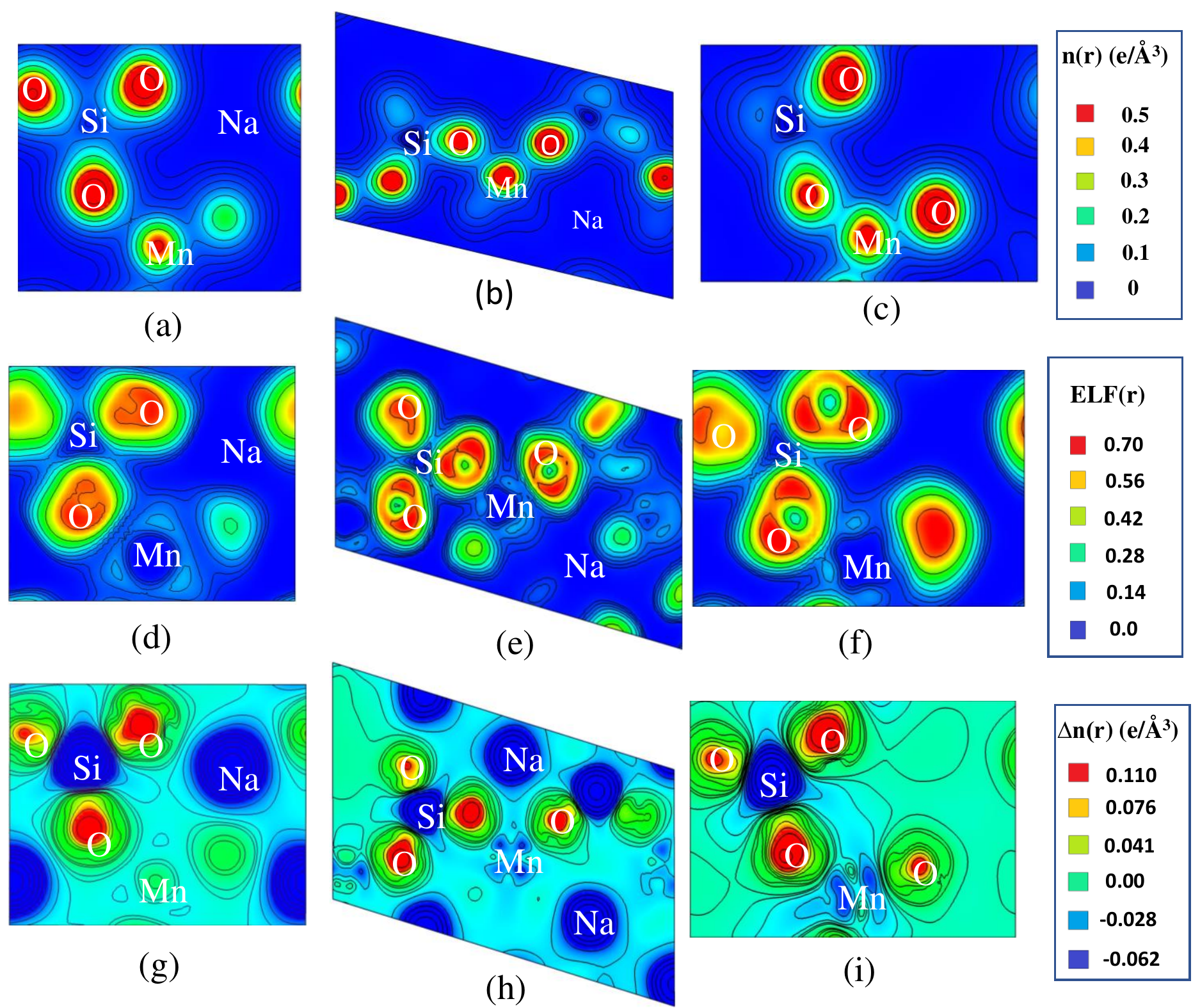}
\centering
\caption{Charge density, electron localization function and charge transfer plots in (a), (d) and (g) for Na\textsubscript{2}MnSiO\textsubscript{4}, (b), (e) and (h) for NaMnSiO\textsubscript{4} and (c), (e) and (i) for MnSiO\textsubscript{4}.}
\label{fig:8}
\end{figure*}

 To extend the study on electronic structure, chemical bonding, and related properties during the desodiation process, we have presented here the charge density, charge transfer, and electron localization function (ELF) analyses for the compounds Na\textsubscript{2}MnSiO\textsubscript{4}, NaMnSiO\textsubscript{4}, and MnSiO\textsubscript{4}. Fig. \ref{fig:8} (a), (b), and (c) shows the charge density distribution of Na\textsubscript{2}MnSiO\textsubscript{4}, NaMnSiO\textsubscript{4}, and MnSiO\textsubscript{4}, respectively in appropriate planes showing the bonding interaction between the constituent atoms. In panels (a), (b), and (c) of  Fig. \ref{fig:8}, the Na atoms are inconspicuous indicating the charge depletion at Na sites. Also, there is no charge distribution between Na and the host lattice indicating that there is a strong ionic interaction between Na and the host lattice. The amount of electrons at the Mn and Si sites are lower than that of the corresponding neutral atoms indicating the presence of ionic bonding between  Mn/ Si with O. However, there exist an anisotropy in charge distribution between Si and O as well as Mn and O indicating the presence of noticeable covalent bonding between them. So, one can conclude that the bonding interaction between Mn$-$O as well as Si$-$O are of iono-covalent character. Also, it can be seen that charge density between Mn and O sites get increased as one go from Na\textsubscript{2}MnSiO\textsubscript{4} to MnSiO\textsubscript{4} indicating the increase in the covalent character of Mn$-$O interaction during the desodiation process. These conclusions are consistent with the chemical bonding between constituents inferred from the DOS analysis.
\par The ELF is a topological tool that is being constructed in such a way that it identifies the presence of paired electrons and associate the lone pair of electrons with a maximum value for ELF as 1. In these systems, the maximum ELF value of around 0.75 is found at the oxygen sites. This indicates that the constituents do not possess a perfect covalent bonding interaction or lone pair of electrons in all these systems considered here.  Due to the presence of covalent bonding between Si and O, there is non-spherical ELF distribution between Si and O in sodiated, desodiated, and the intermediate compounds. Changes in charge density around Mn and O atoms can be seen from the ELF plots of Figs. \ref{fig:8} (d, e, f)  as going from Na\textsubscript{2}MnSiO\textsubscript{4} to MnSiO\textsubscript{4}. On the other hand, no apparent change is observed in the charge density at the Si sites. This indicates the involvement of Mn and O atoms in mediating the redox reaction during desodiation/sodiation process. This topic is discussed in more detail in the following analyses.

\par In order to understand the redistribution of charges during the desodiation process, we have plotted the charge transfer as a function of desodiation as shown in Fig. \ref{fig:8} (g$-$i), obtained by subtracting the charge distribution in the compound by the superposition of electrons in the atomic charges in the lattice grid. In this plot, the positive values are associated with charge gain and negative values are associated with charge depletion during the formation of the solid. It can be seen from the plots of Figs. \ref{fig:8} (g), (h) and (i) that the charge gain is mainly happening at the O sites indicated by the positive charge values, while the Si and Na sites have lost charges indicated by their negative charge values. The charge transfer distribution is isotropic at the Na sites substantiating the presence of ionic bonding between O with both Mn and Si in these compounds. On the other hand, the anisotropic charge transfer distribution between Si and O points to the presence of covalent interaction between these atoms. Slightly negative charge transfer values are seen at the Mn sites as the charge is lost from its sites. Systematic increase in  change transfer from Mn site to O can be visible as going from Na\textsubscript{2}MnSiO\textsubscript{4} to MnSiO\textsubscript{4} suggesting the participation of Mn atoms in the redox reaction. One can also notice that the charge transfer at the Mn sites is significantly low compared to Si or Na sites in  Na\textsubscript{2}MnSiO\textsubscript{4} system as shown in Fig. \ref{fig:8}$-$g. This can be traced back to the covalent interaction between the Mn and O sites. The charge transfer discussed in this part is quantified with the help of  Bader effective charge analysis in the next section.

\begin{figure}[h]
\centering
\includegraphics[scale=0.42]{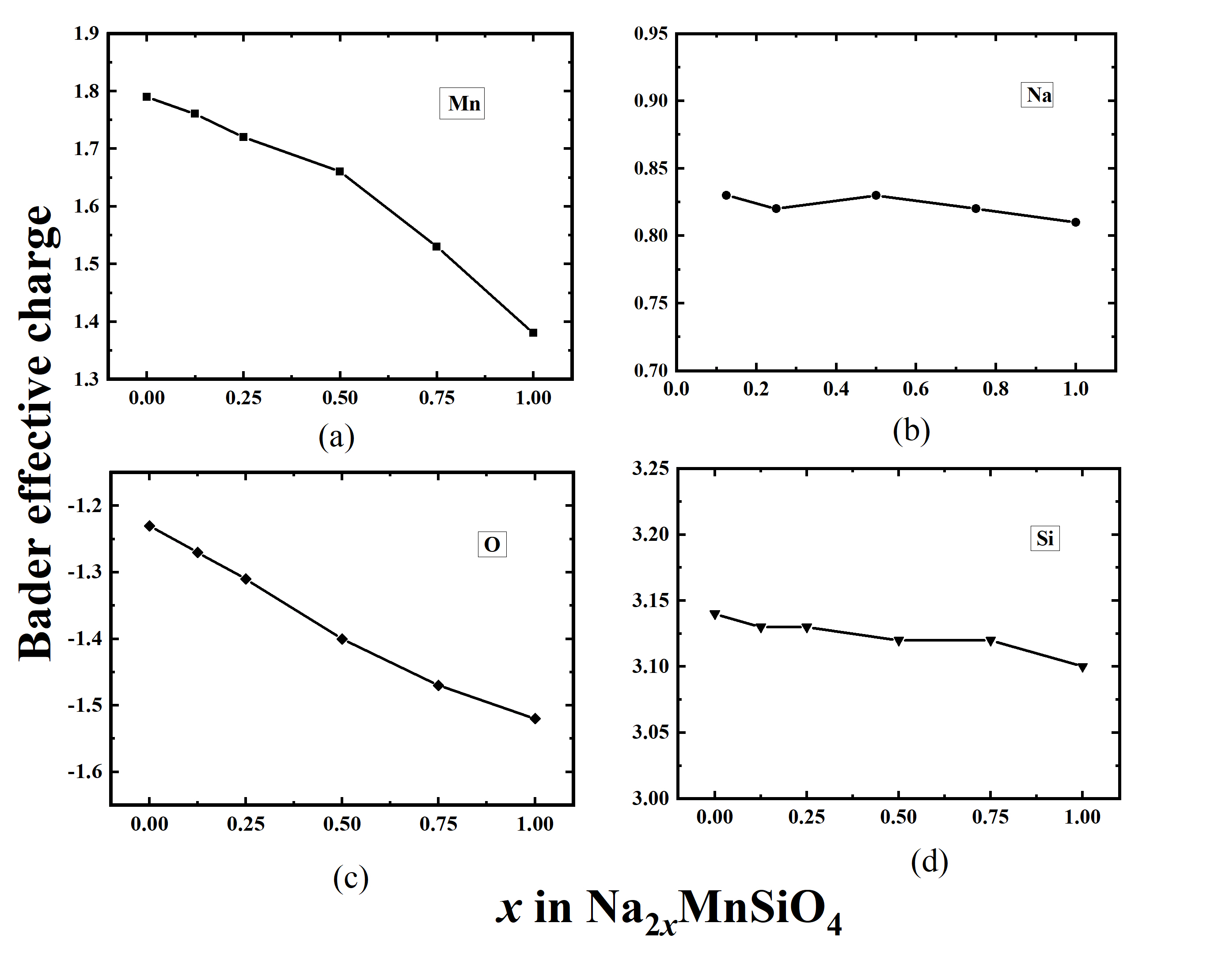}
\caption{Bader net charge of  different atomic sites in Na\textsubscript{2\textit{x}}MnSiO\textsubscript{4} as a function of Na concentration.}
\label{fig:9}
\end{figure}
\par
The Bader method constructs boundaries in each atomic sites based on topological analysis of electronic charge density and thereby defines Bader atomic basin  which is the surface through which the gradient of charge density has zero flux. The volume of these Bader atomic basins serve the purpose of effectively identify the boundaries of individual atoms forming molecules. This method of partitioning charge density enables quantifying the charge associated with each atom. The average Bader effective charges (BEC) associated with each atom during the desodiation process  for  Na\textsubscript{2\textit{x}}MnSiO\textsubscript{4} as a function of Na concentration are shown in Fig: \ref{fig:9}a-d and tabulated in table \ref{table:2}. It can be seen that the BEC at the  Si and Na  sites remain almost the same throughout the desodiation process (see Fig: \ref{fig:9}b and d) and hence, one can conclude that they do not take part in the redox reactions. The BEC value at the Mn and O sites are continuously changing during the desodiation process, as evident from Fig. \ref{fig:9} a and c, indicating their involvement in the redox reactions. The BEC value of Mn changes from 1.78 e in MnSiO\textsubscript{4} to 1.38 e in Na\textsubscript{2}MnSiO\textsubscript{4} (a difference of 0.4 e charge). The charges at the O sites also undergo changes, as can be seen from Fig: \ref{fig:9}a. The BEC value at the O sites change from $-$1.23 to $-$1.52 e when one goes from MnSiO\textsubscript{4} to Na\textsubscript{2}MnSiO\textsubscript{4}(a difference of 0.29 e charge). During the desodiation process, the loss of charge by the removal of Na is compensated by both Mn and O as evident from our BEC analysis. However, though the amount of charge compensation by the Mn appear to be larger than that from O, both these atoms involvement in the charge compensation process can be considered equal due to the fact that the number of O atoms is 4 times greater than the number of Mn atoms in these materials. Usually, in most of the transition metal (TM)-based cathode materials, such a charge compensation is always associated with the TM ions alone. However, apart from the famous Li-excess materials\cite{sathiya2013s, zhao2019stabilizing, yahia2019unified, hy2016performance}, the involvement of oxygen in the redox reaction is found in some polyanionic compounds \cite{zheng2018mechanism}. It may be noted that the Li-counter part of this material (ie. Li\textsubscript{2}MnSiO\textsubscript{4}) also exhibits the oxygen redox reaction but in a much-pronounced manner than in the present system\cite{wang2019high}. Additionally, the charge at both Mn and O sites is less than its classical oxidation states, indicating the presence of covalent interaction between Mn and O atoms.
\begin{table}
\small
  \caption{The calculated Bader effective charges (in units of e) for elements in Na\textsubscript{2x}MnSiO\textsubscript{4} during desodiation.}
  \label{table:2}
  \begin{tabular}{lllll}
    \hline
    System & Mn & Si & O & Na\\
   \hline

     Na\textsubscript{2}MnSiO\textsubscript{4} & 1.38 & 3.10 & -1.52 & 0.81 \\
     Na\textsubscript{1.5}MnSiO\textsubscript{4} &1.53 & 3.12 & -1.47 & 0.82 \\
     NaMnSiO\textsubscript{4} & 1.66 & 3.12 & -1.40 & 0.83 \\
     Na\textsubscript{0.5}MnSiO\textsubscript{4} & 1.72 & 3.13 & -1.31 & 0.82 \\
     Na\textsubscript{0.25}MnSiO\textsubscript{4} & 1.76 & 3.13 & -1.27 & 0.83 \\
     MnSiO\textsubscript{4} & 1.79 & 3.14 & -1.23\\

   \hline
  \end{tabular}
\end{table}

\subsection{Oxygen release analysis} \label{Oxygen release analysis}
\par From the above charge transfer analysis, we have seen that apart from the TM, oxygen has an undeniable part in the electrochemical redox reaction. This might lead to the release of O\textsubscript{2} molecule, as found in the case of several Li-excess cathodes\cite{gao2015selecting, armstrong2006demonstrating, hong2012critical, zheng2019impact}. The oxygen evolution is considered as the leading cause for the irreversible structural changes during the battery operation for such materials.\cite{sharifi2019oxygen} If the Na vacancies are created during the desodiation process, the O atoms may have to donate electrons into the host lattice in order to maintain the charge balance. At some critical stage of this desodiation reaction, the O\textsubscript{2}\textsuperscript{-} is oxidized to O\textsubscript{2}. This, in turn, leads to the instability of the material as the O\textsubscript{2} molecule is released from the material. Considering the involvement of oxygen in the redox process, we examined the possibility of oxygen release by finding the formation energies of O\textsubscript{2} in the material at different concentrations of Na in Na\textsubscript{2x}MnSiO\textsubscript{4}. We have considered the thermodynamic factor alone in analyzing the possibility of O\textsubscript{2} release, even though the O\textsubscript{2} release is dependent on both the kinetic and thermodynamic factors. 
In order to understand the oxygen release reaction we have considered the following reaction.
\begin{equation} \label{eu_eqn8}
Na\textsubscript{2x}MnSiO\textsubscript{4} \longrightarrow Na\textsubscript{2x}MnSiO\textsubscript{4-y} + (y/2)O\textsubscript{2}
\end{equation}
The enthalpy of formation of oxygen molecule for the above reaction is 
\begin{equation}\label{eu_eqn9}
    \Delta H = \frac{E(Na\textsubscript{2x}MnSiO\textsubscript{4-y})+(y/2)E(O\textsubscript{2})-E(Na\textsubscript{2x}MnSiO\textsubscript{4})}{(y/2)}
\end{equation}
Where $E(Na\textsubscript{2x}MnSiO\textsubscript{4})$ is the total energy of the material at Na concentration $x$ and $E(Na\textsubscript{2x}MnSiO\textsubscript{4-y})$ is the total energy of the material with oxygen deficiency. The oxygen deficiency is modeled by removing oxygen atoms with the lowest Bader charge (the more oxidized the oxygen atoms are the higher chances for them to form O$_2$ molecule and leave the structure ) and calculated the total energy by optimizing the structure by supercell approach. The accurate total energy value of the O$_2$ molecule was obtained from the \textit{ab-inito} total energy calculations described by R. Xiao et al.\cite{xiao2012density}. The total energy of the O\textsubscript{2} molecule is found from the more accurate \textit{ab-initio} energies of water and H\textsubscript{2}, as shown below.
\begin{equation}\label{eu_eqn10}
    E(O\textsubscript{2}) = 2E\textsubscript{DFT}(H\textsubscript{2}O) -2E\textsubscript{DFT}(H\textsubscript{2})-2\Delta E\textsubscript{expt}(H\textsubscript{2}O)
\end{equation}
Where E\textsubscript{expt}(H\textsubscript{2}O) is the experimental formation energy of water.
\begin{figure}
\centering
\includegraphics[scale=0.40]{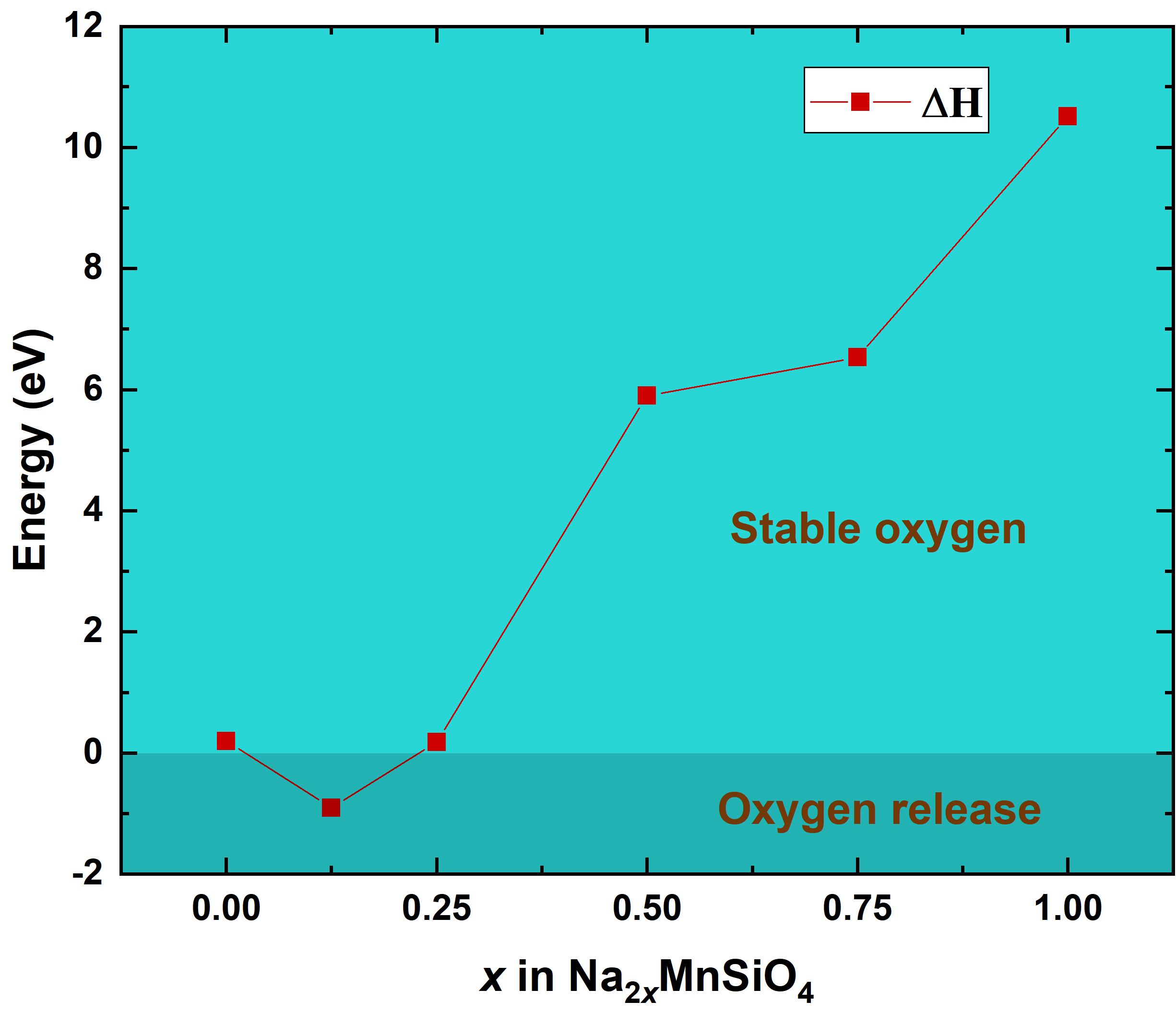}
\caption{The enthalpy of formation of oxygen in Na\textsubscript{2\textit{x}}MnSiO\textsubscript{4} as a function of Na concentration.}
\label{fig:10}
\end{figure}
The enthalpy of formation (\textit{$\Delta$H}) of O\textsubscript{2} for different concentrations of Na in Na\textsubscript{2x}MnSiO\textsubscript{4} is calculated and plotted in Fig: \ref{fig:10}.
The stability of the material is assumed to be inversely related to the release of oxygen. Thus, from this figure, we can observe that this material is not stable as the desodiation progresses beyond 75\% of Na removal. The negative enthalpy of formation of O\textsubscript{2} shows the possibility of instantaneous formation of O\textsubscript{2} and the positive enthalpy of formation indicates that the oxygen release reaction is endothermic and the formation of oxygen molecule is not spontaneous. The enthalpy of formation of oxygen release is slightly positive for Na\textsubscript{0.5}MnSiO\textsubscript{4}. However, when the desodiation process progresses and reaches the composition Na\textsubscript{0.25}MnSiO\textsubscript{4}, the enthalpy of formation  of O\textsubscript{2} is negative, which indicates the possibility of oxygen release. Thus we can conclude that desodiation above 75\% is prone to release of oxygen and the material will not be stable beyond that. If the desodiation process continues beyond 75\% removal of Na then there will be an introduction of holes into the O$-$2\textit{p} states.  This can be clearly seen from the projected DOS shown in  fig \ref{fig:6} that there is an increase in the O$-$2\textit{p} DOS in the conduction band minimum when one desodiate the system with Na removal beyond 75\%.  At this concentration and beyond, the additional Na vacancies introduced in the structure during the desodiation process induce defect levels from which the electrons can transition to the O$-$2\textit{p} unoccupied states since the O-2p states in the conduction band minimum (CBM) shifted to lower energy of around 0.5\,eV when x is smaller than 0.5. The energy gain during such a process will be sufficient to break the Si$-$O bond and lead to irreversible capacity loss.

\subsection{Crystal orbital Hamilton population analysis}
\par
\begin{figure*}[h]
\includegraphics[scale = 0.65]{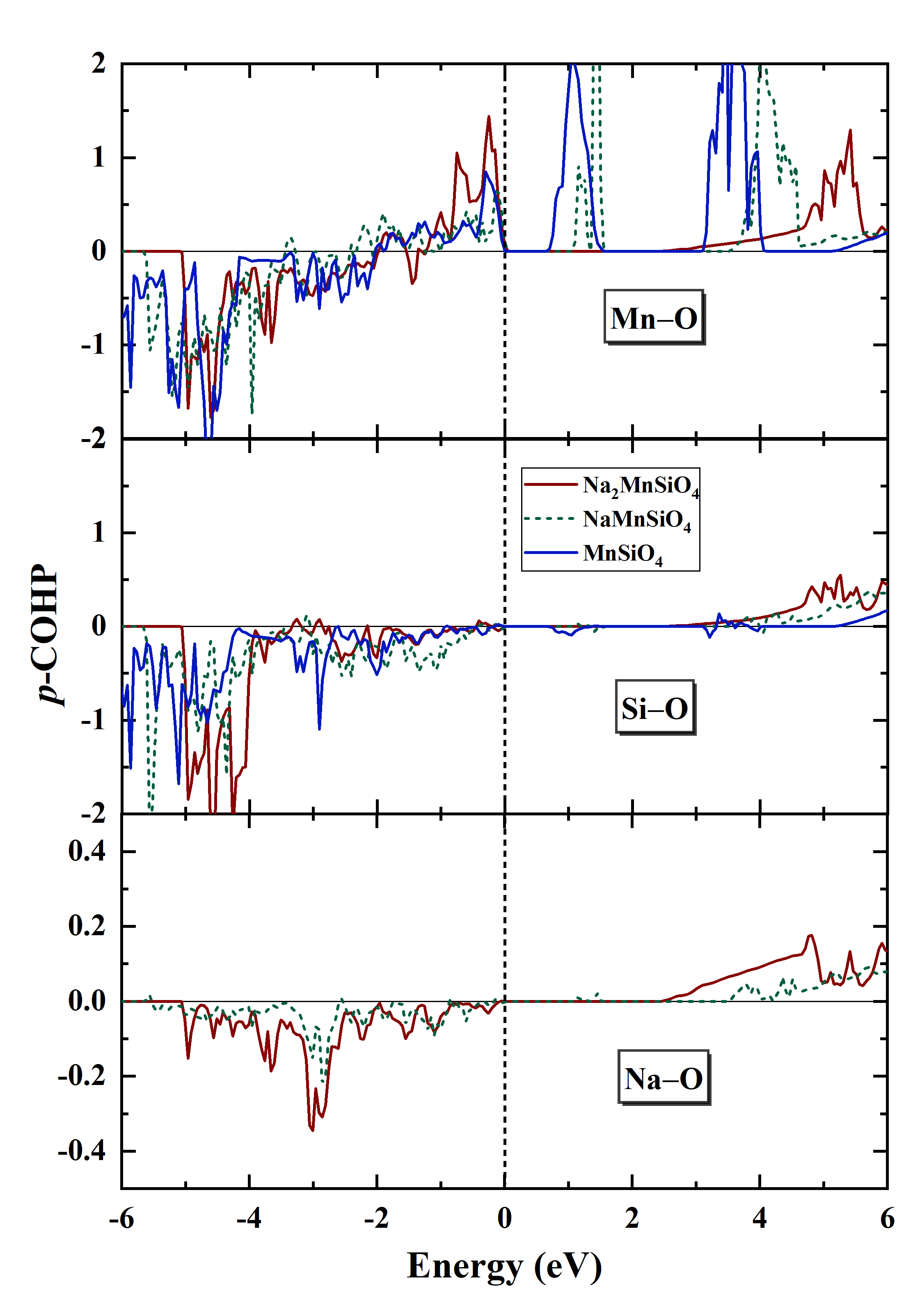}
\centering
\caption{Projected crystal orbital Hamilton population for Mn$-$O, Si$-$O and Na$-$O interactions in Na\textsubscript{2}MnSiO\textsubscript{4}, NaMnSiO\textsubscript{4} and MnSiO\textsubscript{4}. The negative COHP indicates the bonding interaction and positive COHP indicates the antibonding interaction.The Fermi level is set to zero.}
\label{fig:11}
\end{figure*}

 In order to analyze the bonding interaction and strength of chemical bonding between the constituents of Na\textsubscript{2}MnSiO\textsubscript{4} during desodiation, the  crystal orbital Hamiltonian population (COHP) analysis is done for the fully sodiated, completely desodiated, and intermediate compound NaMnSiO\textsubscript{4}. The COHP resolves the density of states into bonding, antibonding, and nonbonding energy regions by partitioning the band structure energy into its orbital-pair interactions.
The negative value of COHP indicates the bonding interaction and the positive value of COHP indicates the antibonding interaction. The maximum filling of bonding states and the empty antibonding states can bring extra stability in to the system. The Mn$-$O pair COHP for MnSiO\textsubscript{4} given in Fig. \ref{fig:11} show that the bonding states are present in the entire valence band region except for the top of the valence band. From -1 eV to VBM, there are noticeable antibonding states present in the Mn-O pairs. The antibonding states at the VBM are the result of the hybridization between Mn$-$\textit{d} and O$-$\textit{p} orbital in the tetrahedral crystal field environment. It may be noted that these antibonding states get decreased by desodiation, as evident from the Fig.\ref{fig:11}. The reduction of antibonding states for the Mn-O hybrid by desodiation strengthen this bond. This can be further substantiated by the integrated COHP (ICOHP) analysis. The ICOHP obtained by the intergration of COHP in the entire VB  gives a direct measure of the bond strength between the corresponding atoms. The smaller value of ICOHP indicates the strong bond strength and the higher value of ICOHP indicates the relatively weak nature of the bond. Due to the strengthening of Mn$-$O bond with desodiation, the ICOHP value up to VBM in the valence band for Mn$-$O reflects the bond strength change from $-$1.68, $-$2.75 and $-$3.34~eV for Na\textsubscript{2}MnSiO\textsubscript{4}, NaMnSiO\textsubscript{4} and MnSiO\textsubscript{4}, respectively. The increase in the bond strength happens due to the removal of electrons from the hybridized Mn$-$O antibonding states as described above. This is also a clear indication of increasing covalency between the Mn-O interaction as the Na atoms leave the structure during desodiation. 
\par Let us now discuss the change in the Si-O bonding interaction during the sodiation process. The analysis of COHP for Si-O bonding hybrid shows that all the bonding states are filled with negligible antibonding states in the entire valence band region in all the three compounds.  This indicate that the bonding interaction between Si$-$O brings extra stability to the system. The desodiation process significantly influences the bonding interaction between Si and O and as a consequence of that, the bonding states are significantly redistributed during the desodiation process as shown in the middle panel of Fig. \ref{fig:11} . However, our ICOHP values for Si-O bond as a function of desodiation are $-2.51$~eV, $-2.56$~eV and $-2.34$~eV for Na\textsubscript{2}MnSiO\textsubscript{4}, NaMnSiO\textsubscript{4} and MnSiO\textsubscript{4}, respectively, which indicates that though desodiation process influence the bonding interaction between Si and O, it does not influence the Si$-$O bond strength signifigantly. The Na$-$O bonding interaction was also analyzed using the COHP as shown in Fig. \ref{fig:11}. From this figure, it is clear that there is noticeable bonding interaction exists between Na and O, which brings sharp bonding states around $-3$~eV in the valence band and this bonding states are almost diminished to half when we go from Na\textsubscript{2}MnSiO\textsubscript{4} to NaMnSiO\textsubscript{4}. From this distinct bonding interaction between Na and O could explain the involvement of O in the redox process. Both Si$-$O and Na$-$O bonding hybrids show that there is negligibly small antibonding states appear in the entire valence band from these bonds and hence these two bonds bring stability to the lattice.
\par In order to understand the role of magnetic ordering on the formation of bonding and antibonding hybrids, we have calculated the COHP for the constituents pair for Na\textsubscript{2}MnSiO\textsubscript{4} in the ferromagnetic as well as antiferromagnetic configuration. Interestingly the antibonding states appeared in the higher energy region of  the valence band in the Mn$-$O hybrid present in both ferromagnetic as well as antiferromagnetic configurations. This indicates that the electronic structure is not altered significantly when one goes from ground state antiferromagnetic configurations to the ferromagnetic configuration. This is consistent with our total energy analysis, which shows that the energy gain from the ferromagnetic configuration to the antiferromagnetic configuration is minimal with a value of 7 meV/f.u. So, for computational simplicity, we have performed the COHP analysis in the ferromagnetic ordering only.  During the desodiation process, as a result of the charge compensation, electrons will be taken from these Mn$-$O antibonding states in the higher energy region of the VB and consequently, part of these antibonding states moved to the CB, forming an IB state. The design of materials with  IB states is of high current interest to develop higher efficiency solar cells based on IB materials. The IB materials are usually obtained by the introduction of defects, transition metal substitutions, and nanoengineering in semi conductors \cite{luque2012understanding, palacios2008transition, soga2006nanostructured}. In this study, we have shown here that the IB semiconductor materials can be designed through electrochemical processes also. The present observation of the formation of IB semiconductors by electrochemical process will motivate experimentalists to identify potential IB semiconductors for high efficiency solar cells.

\begin{figure*}[h]
\includegraphics[scale = 0.25]{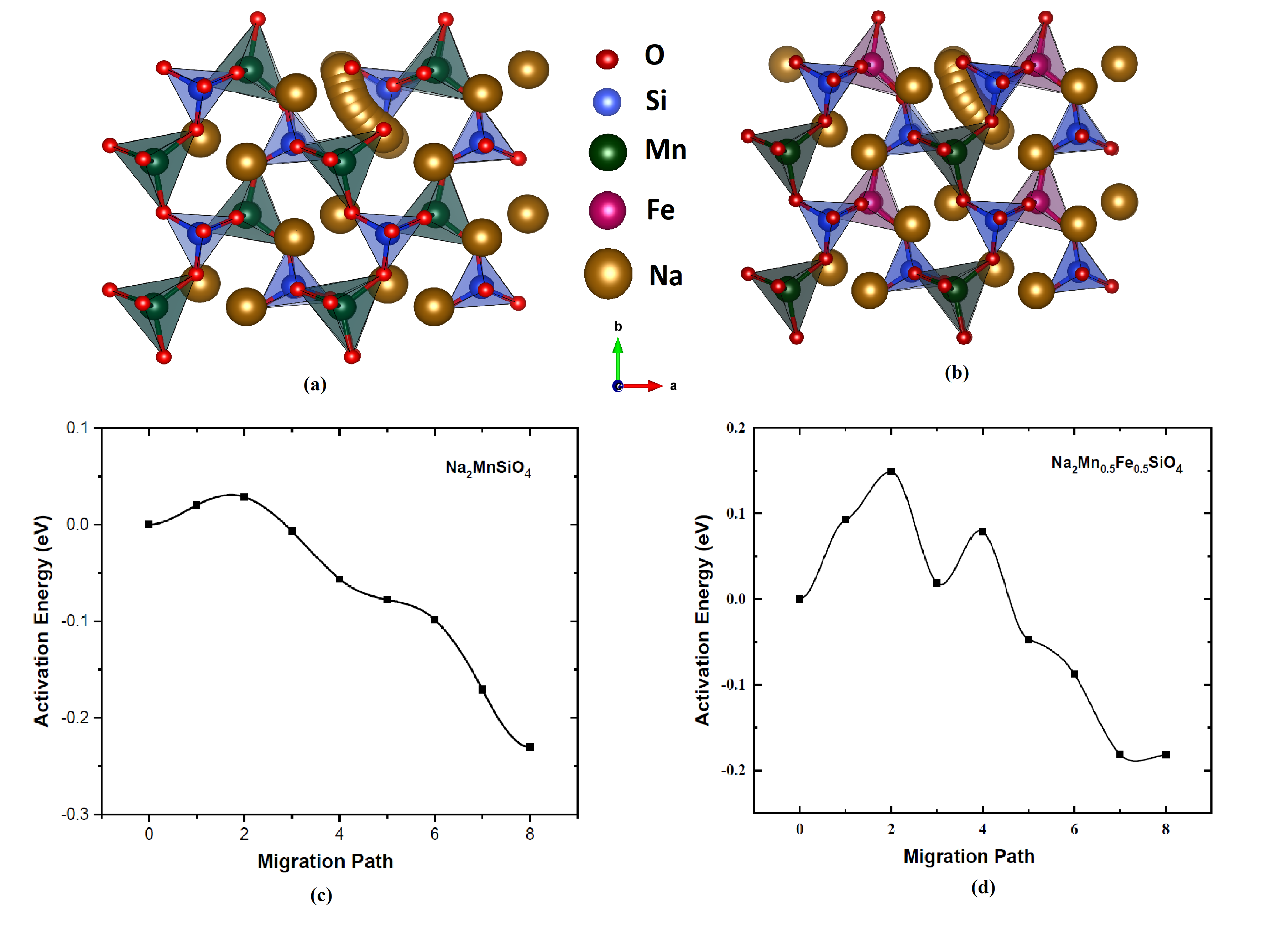}
\centering
\caption{The diffusion path for Na ion in Na\textsubscript{2}MnSiO\textsubscript{4} and Na\textsubscript{2}Mn\textsubscript{0.5}Fe\textsubscript{0.5}SiO\textsubscript{4} are shown in figure a and b respectively and the corresponding activation energy at each points on the diffusion path are shown in figure c and d.}
\label{fig:12}
\end{figure*}

\subsection{Analysis of Na diffusion in Na\textsubscript{2}Mn\textsubscript{x}Fe\textsubscript{1-x}SiO\textsubscript{4}}

\par In order to understand the role of TM substitution on the  diffusion properties of Na in Na\textsubscript{2}MnSiO\textsubscript{4}, we have performed the climbing image-nudged elastic band method (CI-NEB) calculations for Na\textsubscript{2}MnSiO\textsubscript{4} as well as Na\textsubscript{2}Mn\textsubscript{0.5}Fe\textsubscript{0.5}SiO\textsubscript{4}. The parent material Na\textsubscript{2}MnSiO\textsubscript{4} shows a multitude of diffusion paths for different directions \cite{kuganathan2018defects, zhang2015ion}. We have considered the lowest activation energy diffusion path that is along the \textbf{b} direction for the present study and the same path was also adapted for the Fe substituted system to compare the substitutional influences on Na diffusion. The activation energy for Na diffusion in Na\textsubscript{2}MnSiO\textsubscript{4} is found to be 0.26~eV, and this value is comparable with earlier studies on this compound using a similar approach \cite{kuganathan2018defects, zhang2015ion}. Compared to Na\textsubscript{2}MnSiO\textsubscript{4}, Na\textsubscript{2}Mn\textsubscript{0.5}Fe\textsubscript{0.5}SiO\textsubscript{4} has many peaks in the diffusion energy path and the activation energy is also increased to 0.33 eV. The study conducted by Zhang \textit{et al.}\cite{zhang2015ion} revealed that the diffusion barrier of Na in Na\textsubscript{2}MnSiO\textsubscript{4} is much lower in all the hopping paths compared to that for Li in Li\textsubscript{2}MnSiO\textsubscript{4}, and hence the diffusion coefficient is lowered by a factor of 10. In the parent material  Na\textsubscript{2}MnSiO\textsubscript{4} the Na ion along their diffusion paths cross Si, Mn and O atoms, and it is understood that the interaction with Mn\textsuperscript{2+} may have a more significant influence on the migration of Na ions. By analyzing the diffusion path, it can be reasoned that the Na ions favor a path that is far from the interacting Mn\textsuperscript{2+} ions. This is also evident for the Fe substituted system that the Na ions prefer to move away from Fe\textsuperscript{2+} ions. The Fe substitution significantly changes the diffusion path, and many peaks and valleys were appeared along with an increase in the height of the barrier. 
The diffusion coefficient for the corresponding paths are found using the Arrhenius equation 
\begin{equation}\label{eu_eqn11}
 D = d\textsuperscript{2}\nu ~exp\left ( \frac{-E\textsubscript{a}}{k\textsubscript{B}T} \right )
\end{equation}
Where $d$ is the length of the diffusion path, $E_a$ is the activation energy, $k_B$ is the Boltzmann's constant, T is the temperature and $\nu$ is the attempt frequency. $\nu$ = 10\textsuperscript{13} Hz and T =300 K are assumed. The diffusion coefficient calculated for the pristine material is 3.63 $\times$ 10\textsuperscript{-7} cm/s and for the Fe substituted system is 2.47 $\times$ 10\textsuperscript{-8} cm/s. Despite the difference in activation energy, the diffusion coefficient differs only in a single order of magnitude.
\par Our structural optimization based on total energy calculations show that the equilibrium structural parameter and the  volumes are not changed significantly by substitution of Mn with Fe in  Na\textsubscript{2}MnSiO\textsubscript{4} to obtain Na\textsubscript{2}Mn\textsubscript{0.5}Fe\textsubscript{0.5}SiO\textsubscript{4}. Thus the difference in the activation energies can only be attributed to the effect of the increase in the electron count in the system by Fe substitution. It is interesting to note that  Jena. \textit{et al.}\cite{jena2017engineering} have found a correlation between the number of valence electrons in the transition metal and the Li\textsuperscript{+} ion diffusion. From this study, one can conclude that more valence electrons from the transition metals could offer a higher attractive potential to the diffusing of ions. Adapting this theory, and noting that the Fe\textsuperscript{2+} has one extra electron in its \textit{d}$-$states, we could argue that the Fe atom poses an attractive potential to the diffusing Na-ion, which consequently increases the activation barrier. Moreover, the Fe atoms are in close proximity to the diffusing Na ions throughout the path, especially in the initial stages. This may be the cause for the wiggles seen in the activation energy where the combined attractive potential of the Mn and Fe is high.

\section{Conclusions}
First principles based methods are used to study the thermodynamic and electronic properties of the poly anionic compound Na\textsubscript{2}MnSiO\textsubscript{4} for its application as a cathode material in the field of Na-ion battery. This material has a 3D connected tetrahedra of SiO\textsubscript{4} and MnO\textsubscript{4} units with Na atoms in the voids. The existence of a strong SiO\textsubscript{4} tetrahedral network maintains the structural stability of this material during the battery performance. We used a cluster  expansion based method to identify the ground state configurations of Na\textsubscript{2}MnSiO\textsubscript{4} during the desodiation process. Further thermodynamic, as well as electronic structure calculations were performed and the main findings of this study are listed below
\begin{itemize}
  \item The Monte Carlo simulation at finite temperature shows the presence of a plateau region at the end of the desodiation process in the voltage profile. This points to the existence of a two-phase region and as a consequence of that, there is a miscibility gap. The voltage profile also manifest the presence of a solid solution region in between $x$ =  0.1 to $x$ = 0.75 in Na\textsubscript{2x}MnSiO\textsubscript{4} indicating the possibility of ordering between the Na and vacancy at these concentrations.

  \item The DOS analysis shows that IB states are formed during the desodiation process, and hence one would expect drastic changes in the optical properties during desodiation. So, the measurement of the lower energy optical spectra could deduce the amount of desodiation in the system. Usually, the intermediate band states are introduced in semiconductors by introducing defects, transition metal substitution, nanoengineering, etc., and the present study suggests that through electrochemical process also one can introduce the IB states in semiconductors. So, higher efficiency new generation IB solar cells can be obtained from this process. 
  
  \item From various chemical bonding analysis we found that the bonding interaction between Mn and O is of mixed iono-covalent in nature, while that between Si and O is having dominant covalent character. The Na-O bonding interaction shows an expected ionic bonding. From Bader charge and the ELF analysis, we found that the Mn and O atoms undergo changes in their charge density during the desodiation process indicating their involvement in the redox process. Also, from the oxygen evolution reaction study, we identified that when the Na removal is beyond 75\% during desodiation, there is a high possibility of  oxygen evolution in this system.  
  \item The COHP analysis identifies that the higher energy valence band region is made up of the antibonding states of Mn-O bonding hybrid. These antibonding states get depleted and move to the CBM as the electrons will be taken from these states during the desodiation process. The integrated COHP values suggest the increasing bond strength between Mn and O as the desodiation progresses. Our COHP analysis also shows that the bonding states for the Na$-$O bonding hybrid almost halved during the dessodiation as going from Na\textsubscript{2}MnSiO\textsubscript{4} to NaMnSiO\textsubscript{4} and this further confirms the anionic reduction mechanism present in the redox process of this system.
  
  \item The Na diffusion studies are performed for Na\textsubscript{2}MnSiO\textsubscript{4}  and the partially Fe substituted Na\textsubscript{2}Mn\textsubscript{0.5}Fe\textsubscript{0.5}SiO\textsubscript{4} to understand the influence of transition metal substitution in the diffusion of Na through this material. The partial substitution of Fe increases the activation energy and introduces several wiggles to the activation energy curve. This leads to the inference that the introduction of an extra valence electron to the system changes the net interaction of Na with the surrounding environment and thereby changes the activation barrier.
\end{itemize}

These electrochemical and electronic structure findings on Na\textsubscript{2}MnSiO\textsubscript{4} reconfirms the suitability of this material for Na-ion battery applications.

\begin{suppinfo}

The following file is available free of charge.
\begin{itemize}
  \item Supporting information: The optimized structural parameters of Na\textsubscript{2}MnSiO\textsubscript{4}, NaMnSiO\textsubscript{4} and MnSiO\textsubscript{4}. Comparison between the structural parameters of Na\textsubscript{2}MnSiO\textsubscript{4} and Na\textsubscript{2}Mn\textsubscript{0.5}Fe\textsubscript{0.5}SiO\textsubscript{4}
  
\end{itemize}

\end{suppinfo}

\begin{acknowledgement}

 The authors are thankful for the helpful comments from Julija Vinckeviciute and Sanjeev Kolli, Materials Department, University of California, Santa Barbara, regarding CASM. Computer resources are provided by SCANMAT center, Central University of Tamil Nadu.

\end{acknowledgement}


\bibliography{manuscript_vishnu}

\end{document}